\documentclass[useAMS,usenatbib]{mn2e}
\usepackage{epsfig,aastex_hack,amssymb,longtable}
\newcommand{\kms}{\, {\rm km~s}^{-1}}

\newcommand{\kpch}{\>h^{-1} {\rm kpc}}

\newcommand{\Msolh}{\>h^{-1} M_{\odot}}

\def\gsim { \lower .75ex \hbox{$\sim$} \llap{\raise .27ex \hbox{$>$}} }
\def\lsim { \lower .75ex \hbox{$\sim$} \llap{\raise .27ex \hbox{$<$}} }
\def\epsfac {1} 
\def\epsfaca {1.05} 

\title[Hiding Cusps in Cores]{Hiding Cusps in Cores: Kinematics of Disk Galaxies
 in Triaxial Dark Matter Halos}

\author[E. Hayashi \& J.F. Navarro]{Eric Hayashi$^{1,\star}$ 
and Julio F. Navarro$^{1,2,3}$\\
$^1$Max Planck Institute for Astrophysics, Karl-Schwarzschild Strasse 1,
Garching, Munich, D-85740, Germany \\
$^2$Department of Physics and Astronomy, University of Victoria, 
 Victoria, BC V8P 1A1, Canada\\
$^3$Fellow of CIAR and of the J.S.Guggenheim Memorial Foundation\\
$^\star$E-mail:ehayashi@mpa-garching.mpg.de}

\begin{document}
\maketitle

\begin{abstract}
We study the kinematics of gaseous disks in triaxial dark matter halos using
the closed-loop orbit solutions in non-axisymmetric potentials. The orbits are
in general non-circular and, for given triaxiality, their ellipticity depends
on the ratio of escape to circular velocities, $V_{\rm esc}^2/V_c^2$. This
ratio increases steeply towards the center for cold dark matter (CDM) halo
density profiles, implying that even minor deviations from spherical
symmetry may induce large deviations from circular orbits in the velocity
field of a gaseous disk, especially near the center. This result suggests that
caution should be exercised when interpreting constraints on the presence of
density cusps in the dark halo derived from the innermost velocity
profile. Simulated long-slit rotation curves vary greatly in shape, depending
primarily on the viewing angle of the disk and on its orientation relative to
the principal axes of the potential. ``Solid-body'' rotation
curves---typically interpreted as a signature of a constant density core in
the dark matter distribution---are often obtained when the slit samples
velocities near the major axis of the closed loop orbits. Triaxial potentials
imprint specific symmetries in 2D velocity fields, generally inducing
``twists'' in the isovelocity contours and anti-symmetric patterns in opposite
quadrants. We suggest that triaxial halos may be responsible for the variety
of shapes of long-slit rotation curves of low surface brightness (LSB)
galaxies, as well as for the complex central kinematics of LSBs, which are
sometimes ascribed to the presence of ``radial motions'' in the gas. We argue
that LSB rotation curves might be reconciled with the structure of CDM halos
once the effects of halo triaxiality on the dynamics of gaseous disks are
properly taken into account.
\end{abstract}


\section{Introduction}
\label{sec:intro}

It is commonly believed that the inner regions of low surface
brightness (LSB) galaxies are ideal probes of the inner structure of
dark matter halos.  Although estimates of the stellar
mass-to-light ratio remain somewhat uncertain, the baryonic component
contribution to the mass budget in these galaxies is generally thought
to be small.  Under this assumption, dynamical tracers of the
potential such as rotation curves are expected to cleanly trace the
dark matter distribution in LSBs. This provides important
astrophysical clues to the nature of dark matter, since the spatial
distribution of dark material in these highly non-linear regions is
expected to be quite sensitive to the physical properties of the dark
matter.

LSB rotation curves may thus be contrasted directly with theoretical
predictions of the inner structure of halos, and there is now an
extensive body of work in the literature that reports substantial
disagreement between the shape of LSB rotation curves and the circular
velocity curves of simulated cold dark matter (CDM) halos \citep[see,
e.g.,][]{FLORES94,MOORE94,MCGAUGH98,DEBLOK01}.  Some of these rotation
curves are fit better by circular velocity curves arising from density
profiles with a constant density ``core'' rather than by the ``cuspy''
density profiles commonly used to fit the structure of CDM halos
\citep[hereafter NFW]{NFW96, NFW97}. This discrepancy adds to a
growing list of concerns regarding the consistency of CDM with
observational constraints on the scale of individual galaxies
\citep[see, e.g.,][]{SELLWOOD00} that has prompted calls for a radical
revision of the CDM paradigm on small scales \citep[see,
e.g.,][]{SS00}.

Before accepting the need for radical modifications to CDM it is important to
note a number of caveats that apply to the LSB rotation curve problem.  For
instance, many of the early rotation curves where the disagreement was noted
were significantly affected by beam smearing in the HI data \citep{SWATERS00}.
The observational situation has now improved substantially thanks to
higher-resolution rotation curves obtained from long-slit and two-dimensional
H$\alpha$ observations \citep[see,
e.g.,][]{DEBLOK02,SWATERS03,SIMON03,SIMON05}. We shall restrict our analysis to
these newer datasets in what follows.

We also note that, strictly speaking, the observational disagreement
is with the fitting formulae used to parameterize the structure of
simulated CDM halos (usually the profile proposed by NFW), rather than
with the structure of simulated halos themselves. Although the fitting
formulae provide a simple and reasonably accurate description of the
mass profile of CDM halos, the radial range over which they have been
validated often does not coincide with the scales where the
disagreement has been identified. This is important, since small but
significant deviations between the NFW profile and simulated halos
have been reported as the mass and spatial resolution of the
simulations has increased
\citep{MOORE98,GHIGNA00,FUK97,FUK01}.  In particular, the central slope of the
density profile has been the subject of much debate.  

The NFW density profile is given by $\rho(r) \propto (r/r_s)^{-1}
(1+r/r_s)^{-2}$, where $r_s$ is a scale radius. This profile
approaches a power-law $\rho(r) \propto r^{-1}$ ``cusp'' near the
center. \cite{MOORE98} and \cite{GHIGNA00} argued that halos simulated
with higher resolution have steeper central density profiles, $\rho(r)
\propto r^{-1.5}$.  However, \cite{HAYASHI04} and \cite{NAVARRO04}
show that the density profile of halos becomes shallower towards the
center and in fact may not actually converge to a well-defined
asymptotic inner power law.  At present, the simulation results of all
groups are in good agreement, and the logarithmic slope of the density
profile is $\simeq -1.2$ or shallower at the innermost resolved radius
\citep{DIEMAND04}.  Although there is no broad consensus yet regarding
how deviations in simulated halo profiles from fitting formulae like
NFW may affect the comparison with observed rotation curves
\citep[see, e.g.,][]{HAYASHI04}, the fact that the deviations worsen
towards the centre, in addition to the observed scatter in halo
profiles, advise against using extrapolations of simple fitting
formulae such as the NFW profile to assess consistency with
observation.

Finally, it must be emphasized that the ``cusp vs. core'' problem arises when
comparing rotation speeds of gas in LSB disks to spherically-averaged circular
velocities of dark matter halos.  Given that CDM halos are expected to be
significantly non-spherical, some differences between the two are to be
expected.  Previous studies have shown that simulated halos are triaxial objects
with shapes ranging from oblate to prolate
\citep{DAVIS85,BARNES87,FRENK88,WARREN92,JING95,THOMAS98,JING02}.  The angular
momentum of the halo tends to be aligned with the minor axis,
therefore one expects galactic disks to lie in a plane close to the
plane defined by the intermediate and major axes of the halo.  It is
therefore important to take into account deviations from spherical
symmetry in the structure of CDM halos in order to make predictions
regarding the rotation curves of gaseous disks that may be compared
directly to observation.  In fact, asymmetries in the velocity fields
of gas disks have been observed \citep[e.g.,]{COURTEAU03}, although in
normal galaxies they are usually attributed to non-circular motion
induced by bars or spiral arms.

We address this issue here by exploring the closed loop orbits within
simple non-axisymmetric potentials whose mass profiles are consistent
with cold dark matter halos.  We follow the formalism developed by
\cite{GERHARD86} and
\cite{BT87} in the context of triaxial bulges and bar potentials, and by
\cite{FRANX94} and \cite{SCHOEN97} to probe non-axisymmetry in spiral galaxies.
We focus our analysis on the shape and inner slope of the rotation curves
inferred for disks in triaxial halos by simulating long-slit observations of
their velocity fields.  We find that velocities along the slit may deviate
significantly from the circular velocity, a result that may account for the
variety of LSB rotation curve shapes and that may reconcile the ``cores''
inferred for some LSBs with the cuspy mass profile of CDM halos.

\section{LSB rotation curves}
\label{sec:lsbrc} 

\begin{figure} 
\epsscale{\epsfac}
\plotone{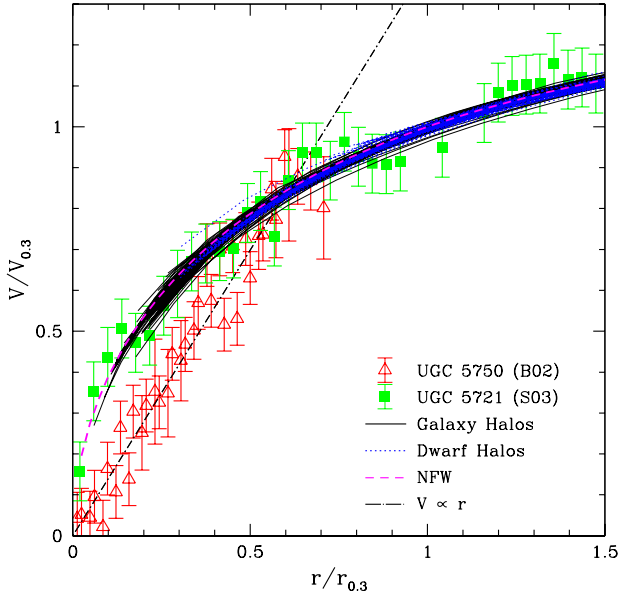}
\caption
[Rotation curves of LSB galaxies and circular velocity profiles of simulated
halos] {Rotation curves of two LSB galaxies from the samples of
\citet[][B02]{DEBLOK02} and \citet[][S03]{SWATERS03}, chosen to illustrate
their various shapes. Rotation curves have been scaled to the radius $r_{0.3}$
and corresponding velocity $V_{0.3}$ (see text for details).  The NFW profile
(dashed line) and the $V_c$ profiles of simulated dwarf- (dotted lines) and
galaxy-sized halos (solid lines) match reasonably well systems like
UGC~5721 but cannot account for those with a linear rise in
velocity with radius, like UGC~5750.\label{figs:rc_allgamma}}
\end{figure}

Figure~\ref{figs:rc_allgamma} illustrates the ``cusp vs core'' disagreement
alluded to above. This figure shows the H$\alpha$ rotation curves of two LSB
galaxies (points with error bars) selected from the sample of
\citet[][B02]{DEBLOK02} and \citet[][S03]{SWATERS03}.  Also shown are the
spherically-averaged $V_c$ profiles of all galaxy-sized cold dark
matter halos presented in \cite{HAYASHI04} and \cite{NAVARRO04}.  In
order to emphasize discrepancies in shape, the rotation curves in
Figure~\ref{figs:rc_allgamma} have been scaled to the radius,
$r_{0.3}$, and velocity, $V_{0.3}$, where the logarithmic slope of the
curve is $d\log V/d\log r=0.3$.  The choice of this scaling is
motivated by the asymptotic slope of the NFW $V_c$ profile, $d\log
V/d\log r=0.5$. In this case, the scaling radius $r_{0.3}=0.411~r_s$
is in the region well resolved by numerical simulation; it is also
easily identified in observed rotation curves, as it lies halfway
between the linear ($d\log V/d\log r=1$) ``core'' region and the
outer, flat ($d\log V/d\log r=0$) rotation curve portion seen in most
disk galaxy rotation curves.

The dashed line (which runs through the simulation data) shows the
$V_c$ profile of an NFW halo, which is fixed and independent of
concentration in these scaled units. We find that the shapes of the
dark halo $V_c$ curves are in fact quite similar to NFW, although they
span a large range of halo masses and have been chosen without regard
to the dynamical state of the halo.

LSB rotation curves, on the other hand, exhibit a wide range of shapes, and the
two galaxies shown in Figure~\ref{figs:rc_allgamma} have been chosen to
illustrate some of the extreme cases in the B02 and S03 datasets. These have
been scaled to $r_{0.3}$ and $V_{0.3}$ as determined by fits with the rotation
curve fitting formula of \cite{COURTEAU97} \citep[see][for details]{HAYASHI04}.
Galaxies like UGC~5721 are roughly consistent with NFW and feature a gentle
turnover between the rising and flat parts of their velocity curves.  Others are
better approximated by a linear scaling of velocity with radius, as expected in
the presence of a constant density core.  One such example, UGC~5750, is shown
in Figure~\ref{figs:rc_allgamma}, and is well-matched by the dot-dashed $V
\propto r$ line.

\section{Closed Loop Orbits}
\label{sec:closedloop} 

Does this result imply that a cusp in the dark matter density profile cannot
be present in galaxies such as UGC~5750? As noted in \S~\ref{sec:intro},
before concluding so one must take into account possible systematic
differences between rotation speed and circular velocity in gaseous disks
embedded within realistic, triaxial halos. This is a complex issue that
involves a number of parameters, such as the degree of triaxiality, the role
of the disk's self-gravity, size, and orientation, as well as the possibility
of transient deviations from equilibrium. 

We explore a simplified scenario here, where the orientation of the principal
axes of the potential is constant with radius and a massless, filled gaseous
disk is placed on one of the symmetry planes of the halo. The velocity field
of the disk may then be examined by solving for the closed loop orbits within
a two-dimensional non-axisymmetric potential. This is analogous to orbits
within a barred potential where the pattern speed of the bar is set to zero,
and has been studied by a number of authors using the epicycle approximation
for the case of weak bars. Following the formalism presented in
\cite{SCHOEN97}, we may write the potential as,

\begin{equation}
\Phi(R,\phi) = \Phi_0(R) + \Phi_m(R) \cos(m \phi_0 - \phi_m(R)),
\label{eq:potpert} 
\end{equation}

where $\Phi_0(R)$ is the unperturbed potential, $\Phi_m(R)$ is a
stationary perturbation to that potential, $\phi_0$ is the azimuthal
angle, and $\phi_m(R)$ is the radially-dependent phase of the
perturbation.  For the simple case we consider here, we set $m=2$ and
require the phase of the (stationary) perturbation to be constant with
radius, $\phi_m(R) = \phi_m$.  We then choose the origin so that
$\phi_m = 0$ along the major (minor) axis of the perturbed potential
for positive (negative) $\Phi_m(R)$.

The closed-loop orbit solutions can be calculated
analytically using the epicyclic approximation and are given in parametric form
by

\begin{eqnarray}
R_{\rm orb} & = & R_0 \left( 1-\frac{a_{1m}}{2} \cos(m \phi_0)\right)\\
\phi_{\rm orb} & = & \phi_0 + \frac{a_{1m} + a_{3m}}{2m} \sin(m \phi_0)\\
V_R & = & m \, V_c(R_0) \, \frac{a_{1m}}{2} \sin(m \phi_0) \\
V_\phi & = & V_c(R_0)\left( 1 + \frac{a_{3m}}{2} \cos(m \phi_0)\right),
\label{eq:vphi} 
\end{eqnarray}

where $(R_0,\phi_0)$ is the guiding centre of the closed loop orbit; $\phi_0 =
\Omega_0 t$; and the (unperturbed) angular and circular velocities are given
by $\Omega_0^2 = {1/R_0~d \Phi/d R}$ and by $V_c(R_0) = \Omega_0 R_0$.

\setcounter{footnote}{0}

\subsection{Case I: A Uniform Perturbation}
\label{sec:unipert} 

In general, the terms $a_{1m}$ and $a_{3m}$
\footnote{$a_{2m}$ and $a_{4m}$ are zero for stationary perturbations} 
are functions of $\Phi_0(R)$, $\Phi_m(R)$ and their first and second
derivatives with respect to $R$.  However, we begin by examining the
simpler case of a uniform perturbation, i.e., $\Phi_m(R) = f
\Phi_0(R)$, where $f$ is a constant.   In this case, these terms in
the closed-loop orbit solutions simplify to:

\begin{eqnarray}
\label{eq:a1m} 
a_{1m} & = & {2 f \over g(R)} \left ( 1 - \frac{V_{\rm esc}^2}{V_c^2} \right) \\
\label{eq:a3m} 
a_{3m} & = & {2 f \over g(R)} \left ( 1 - \frac{5}{2} \frac{V_{\rm esc}^2}{V_c^2} \right)\\
\label{eq:gr} 
g(R) & = & {\Phi_0''/\Omega_0^2-1}, 
\end{eqnarray}

where the escape velocity is $V_{\rm esc} = \sqrt{2 |\Phi_0|}$;
$\Phi_0'' \equiv d^2 \Phi_0/d R^2$; and all terms in
eqs.~\ref{eq:a1m}-\ref{eq:gr} are evaluated at $R=R_0$.  We note that
this perturbation tends to a finite value as $R \rightarrow 0$ and
therefore can result in unphysical solutions to Poisson's equation for
the density distribution at small radii.  However, we explore this
simplified case in order to illustrate the general character of the
closed orbit solutions.  A more realistic perturbation in which $f$
tends to zero as $R \rightarrow 0$ is presented later in this section.

For a power-law potential, $V_c \propto R^{\alpha}$, the function $g$ is
independent of radius, $g=2(\alpha -1)$. Near the center of an NFW profile
$\alpha \approx 1/2$, and the ellipticity of the orbits scales simply as
$\epsilon_{\rm R}\equiv (R_{\rm max}-R_{\rm min})/(R_{\rm max}+R_{\rm min}) =
a_{1m}/2 = f \, (V_{\rm esc}^2/V_c^2-1)$. Note that the shape of the orbit at
some radius depends on the mass profile both inside (through $V_c$) and
outside (through $V_{\rm esc}$) the orbit, since the gravitational forces of
the outlying material do not cancel out in a non-spherical
potential. Quantitatively, the deviation from circularity depends on the
competition between the monopole of the unperturbed potential and the
quadrupole of the perturbation. The larger the central mass concentration the
larger the monopole and the smaller the ellipticity of the orbit for given
fractional perturbation $f$. Indeed, deviations from circular motion are
minimum for a point-mass (Kepler) potential, where $\epsilon_{\rm R}= f$. For
more general mass profiles, such as those of CDM halos, where $V_{\rm esc}$ is
finite at the center and $V_c$ tends to zero, the ellipticity of closed loop
orbits is expected to increase dramatically toward the center, even for
relatively minor deviations from spherical symmetry.

\begin{figure} 
\epsscale{\epsfac}
\plotone{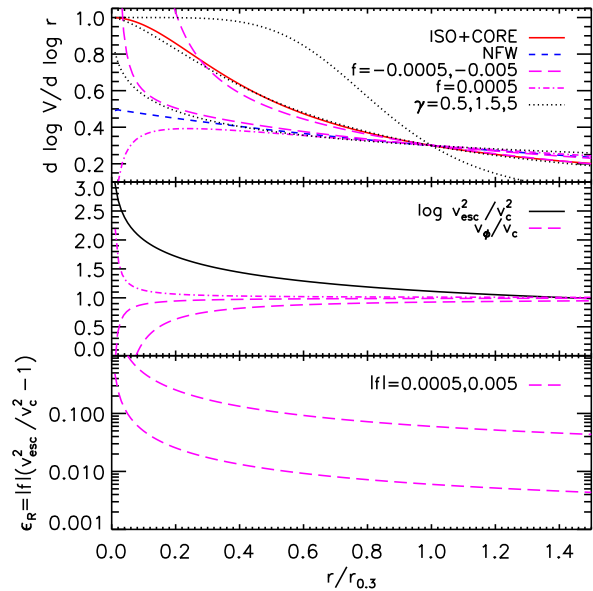}
\caption[log slope] {{\it Top panel:} Logarithmic slope, $d \log V/d \log r$, of
the pseudo-isothermal (solid) and NFW (dashed) circular velocity profiles.
\cite{COURTEAU97} velocity profiles for three values of the parameter $\gamma$
(increasing from left to right) are shown by the dotted curves.  All curves are
normalized to the radius $r_{0.3}$ where $d\log V/d\log r = 0.3$. Long-dashed,
$f<0$, (dot-dashed, $f > 0$) curves correspond to the tangential velocity
profile, $V_\phi$, measured along the major (minor) axis of the closed-loop
orbits of a disk in an NFW potential with an $m=2$ perturbation. The magnitude
of the perturbation is a constant fraction, $f$, of the NFW potential at all
radii. Curves for two perturbation magnitudes are shown, with $|f|$ increasing
from left to right for the long-dashed curves.  Even for rather small
perturbations to an NFW potential, the tangential velocity profile exhibits a
wide variety of shapes, matching the pseudo-isothermal profile over a wide range
of radius, as well as the Courteau profile with various values of the shape
parameter $\gamma$.  {\it Middle panel:} Ratio of $V_\phi$ to the NFW circular
velocity $V_c$ for the same two values of $f$ as in the top panel.  The
departure from circularity depends on the ratio between escape and circular
velocities (shown by the upper curve) The escape velocity becomes much larger
than the circular velocity towards the center of the NFW potential, resulting in
large deviations from circularity even for small perturbations in the
potential. {\it Bottom panel:} Ellipticity of the orbits, $\epsilon_R$, for
constant perturbation amplitudes shown in the top panel.  The ellipticity is
simply related to the ratio between escape and circular velocities and therefore
increases toward the center of the NFW potential.  Note that the epicyclic
approximation breaks down at small radii for sufficiently large perturbation
magnitudes, resulting in invalid solutions with $\epsilon_R >1$.
\label{figs:dlogv}}
\end{figure}

Along an orbit, the maximum deviations from circular motion occurs along the
minor and major axis of the perturbation. Assuming an NFW profile for the
unperturbed potential, i.e., $\Phi_0 = \Phi_{\rm NFW}$, we consider in
Figure~\ref{figs:dlogv} the radial dependence of the tangential velocity,
$V_\phi$, for various values of $f$.  

The top panel shows the logarithmic slope of the $V_\phi$ velocity profile along
the $\phi_0=0$ axis for $m=2$ perturbation magnitudes of $0.05\%$ and $1\%$,
respectively.  For negative values of $f$, the $\phi_0 = 0$ axis corresponds to
the minor axis of the isopotential contours and the major axis of the closed
loop orbits, and the tangential velocity is at a minimum at this point on the
orbit.  The middle panel of Figure~\ref{figs:dlogv} shows the ratio
$V_\phi/V_c$, and 
the bottom panel shows the ellipticity of the orbits, which scales as the ratio
between the escape and circular velocities and therefore increases toward the
center of the NFW potential.

Also shown in the top panel are the logarithmic slopes of the NFW (dashed) and
pseudo-isothermal (solid) velocity profiles, as well as the rotation curve
fitting formula of \cite{COURTEAU97} (dotted). The pseudo-isothermal velocity
profile is given by $V_{\rm iso}^2(r) = V_\infty^2 (1 - ({r_c}/{r})
\tan^{-1}({r}/{r_c}))$, where $V_\infty$ is the asymptotic velocity and $r_c$
is the radius of the constant density core in this model.  The Courteau
velocity profile is given by $V_{\gamma}(r)=V_0
(1+(r/r_t)^{-\gamma})^{-1/\gamma}$, where $V_0$ and $r_t$ are dimensional
scaling parameters and $\gamma$ is a dimensionless parameter that
characterizes the shape of the rotation curve.  This three-parameter formula
provides excellent fits to LSB rotation curves with a wide variety of shapes
\citep{HAYASHI04}.

Figure~\ref{figs:dlogv} shows that even small perturbations in the potential can
result in large changes in the shape of the velocity profile. For $f= -0.005$,
the shape of the $V_\phi$ profile is similar to that of the pseudo-isothermal
profile (and of that corresponding to $\gamma=1.5$) down to radii as small as
one-third of $r_{0.3}$. Reducing the magnitude of $f$ has the effect of moving
inwards the radius (in units of $r_{0.3}$) where the deviations become important
but, because the $V_{\rm esc}^2/V_c^2$ term formally diverges at the center, the
shape of the rotation curve always deviates significantly from the true circular
velocity curve somewhere near the center.  Also shown in Figure~\ref{figs:dlogv}
is the velocity profile along the $\phi_0 = 0$ axis for a perturbation with
$f=0.0005$.  Note that this is equivalent to the velocity profile along the
minor ($\phi_0 = 90^\circ$) axis for $f=-0.0005$ (see eqs.~5 and 7).  Along this
axis, the tangential velocity is at a maximum, and the logarithmic slope of the
inner velocity profile is actually shallower than NFW.  This implies that the
shape of the rotation curve of a disk in a non-spherical potential varies
depending on the viewing angle from which it is observed, a possibility we
investigate further in \ref{sec:simrot}.

We note that the analytic expressions for the closed orbit solutions
are derived assuming $d \phi_{\rm orb}/d t \simeq d \phi_0/d t$,
therefore the solutions break down for large perturbations where
epicyclic theory is not a good approximation to the true solution for
the orbits.  The bottom panel of Figure~\ref{figs:dlogv} shows that
the ellipticity of the orbits predicted by the epicyclic approximation
reaches values of $\epsilon_R >1$ at small radii for sufficiently
large perturbation magnitudes.  In practice the closed orbit solutions
are valid down to some minimum radius where the epicyclic
approximation still holds, therefore we show the solutions only at
radii where $d \phi_{\rm orb}/d t \lsim d \phi_0/d t$ in all
subsequent figures.

The results shown in Fig~\ref{figs:dlogv} suggest that caution should
be exercised when interpreting the constraints on the slope of the
dark halo inner cusp derived from analyzing the innermost region of
long-slit rotation curves \citep{DEBLOK02, DEBLOK05}. These regions
are not just particularly sensitive to observational error
\citep{SWATERS03, SPEKKENS05} but, as the discussion above
illustrates, one would expect on rather general grounds that gas
velocities near the center might deviate substantially from the
circular speed.

\subsection{Case II: A Radially Varying Perturbation}
\label{sec:varypert} 

In general, the magnitude of the perturbation, $f$, is expected to be
a function of radius and it is instructive to consider perturbations
to an NFW potential that may result in a pseudo-isothermal rotation
curve along the major axis of the closed loop orbits. In order to
derive such a perturbation, we cannot use the simplified solutions in
eqs.~\ref{eq:a1m}-\ref{eq:gr}, but instead set $V_\phi = V_{\rm iso}$ and
calculate numerically the function $f_{\rm iso}(R) =
\Phi_m/\Phi_{\rm NFW}$ required to satisfy eq.~\ref{eq:vphi} for
$\phi_0 = 0$.  Note that, by construction, this perturbation vanishes
as $R\rightarrow 0$ since both the pseudo-isothermal and NFW mass
profiles are convergent at small radii.  

We set the parameters of the unperturbed NFW potential to match a
typical LSB galaxy; the mass, virial radius and scale radius are set
to $M_{200} = 10^{11}~\Msolh$, $r_{200} = 75~\kpch$, and $r_s =
6.25~\kpch$, respectively, giving a peak circular velocity of $V_{\rm
max} = 95.2~\kms$.  The pseudo-isothermal velocity profile is
normalized to match the NFW profile at $(r_{0.3},
V_{0.3})=(2.57~\kpch, 73.5~\kms)$ such that $r_c= 0.95~\kpch$ and
$V_\infty= 99.2~\kms$.  The maximum deviation between the NFW and the
pseudo-isothermal profile within $r < 20~\kpch$ is $11.7~\kms$ at a
radius of $r=0.22~\kpch$.  We emphasize that, although this procedure
specifies a particular radial dependence for $f(R)$, in general {\it
most} well-behaved perturbations will lead to deviations from the
circular speed which become more pronounced toward the center,
generally mimicking the presence of a ``core'' along the $\phi_0=0$
axis.

\begin{figure} 
\epsscale{\epsfac}
\centerline{\epsfig{figure=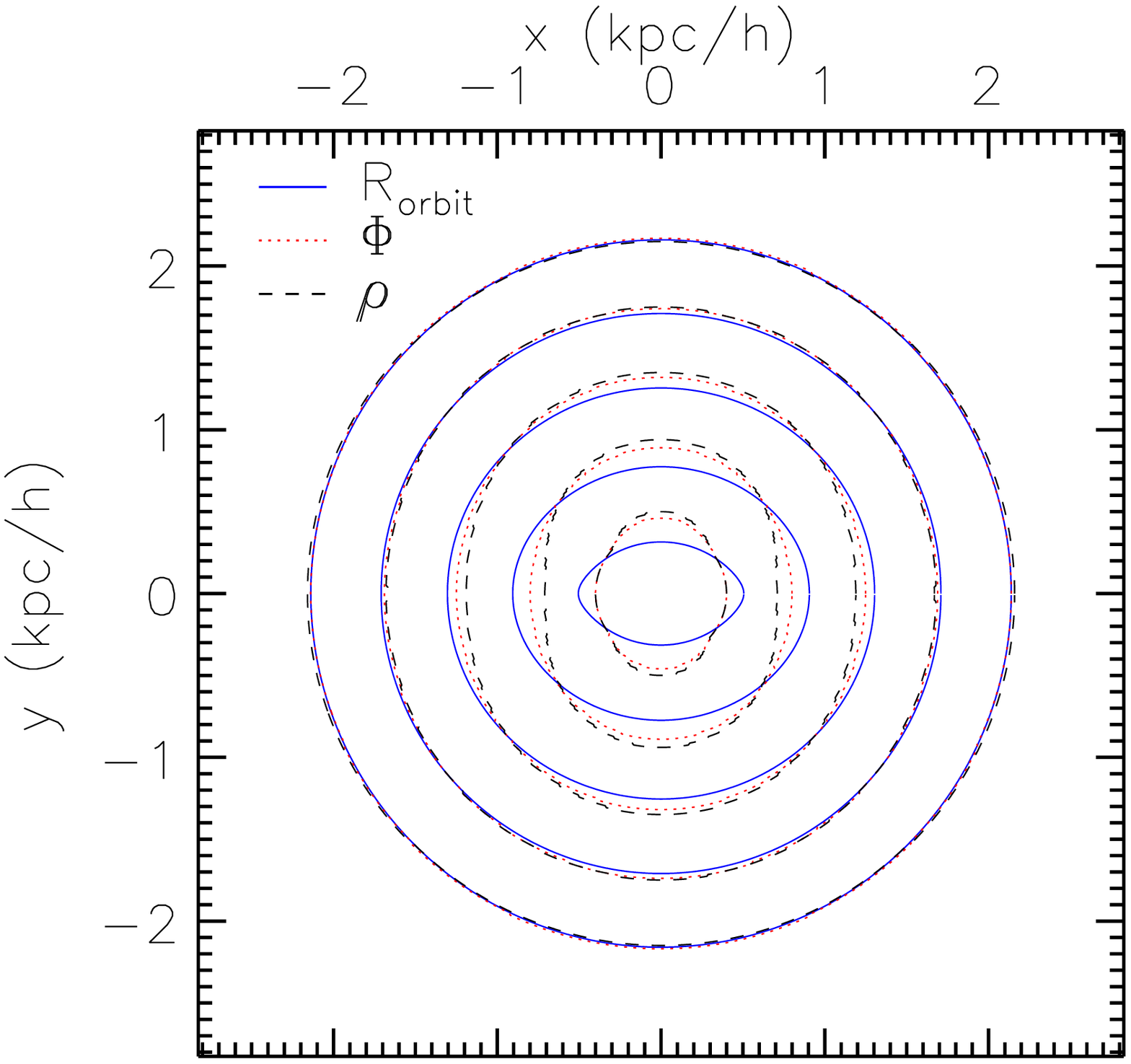,width=\epsfac\linewidth}}
\centerline{\epsfig{figure=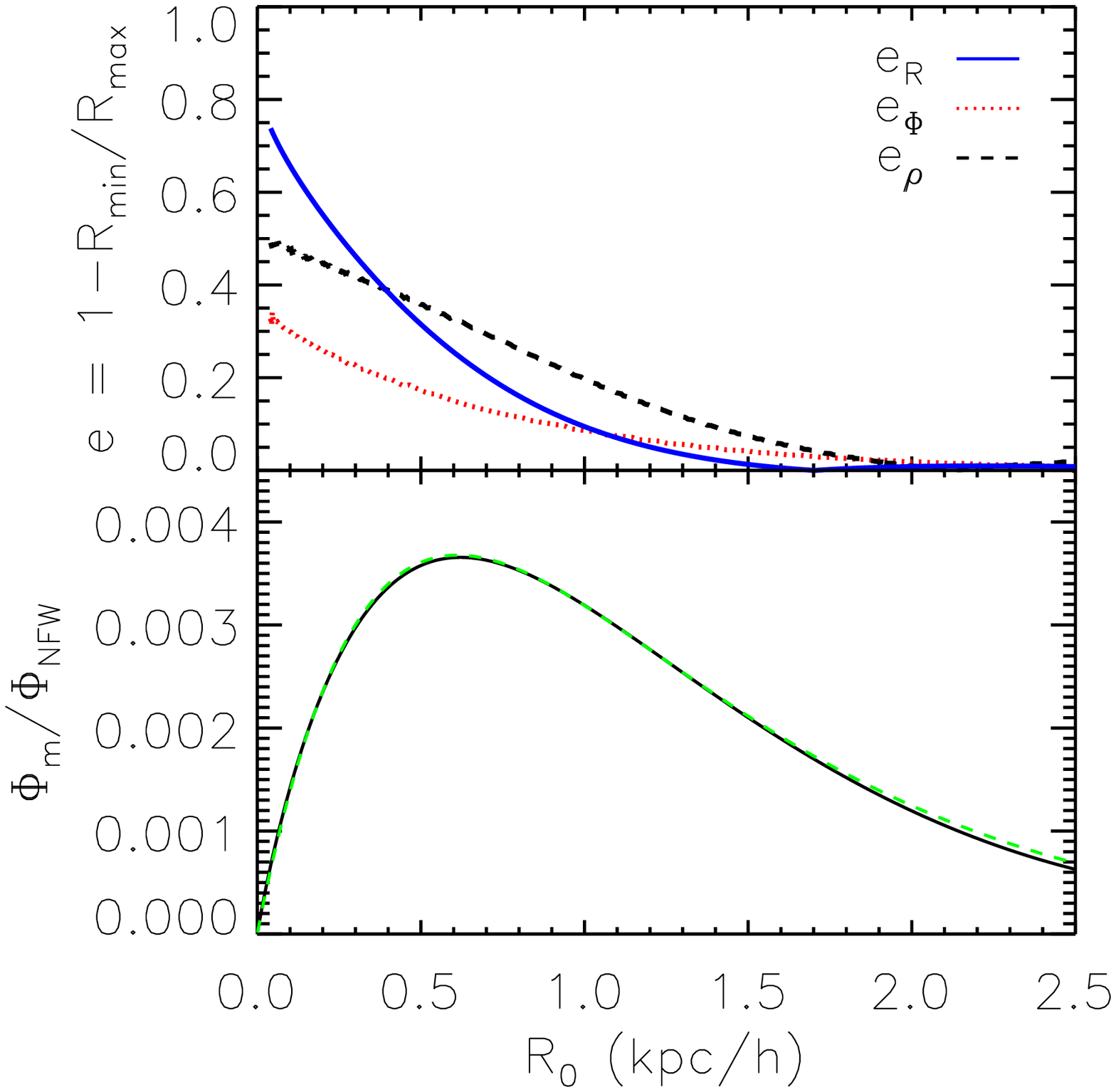,width=\epsfaca\linewidth}}
\caption[contours] {{\it Top panel:} Closed loop orbit trajectories in a
  perturbed NFW potential, as well as isopotential and isodensity contours.
  Orbits are elongated in the direction perpendicular to the major axis of the
  isopotential contours.  {\it Middle panel:} Eccentricity of the orbits, as
  well as that of isopotential and isodensity contours, as a function of the
  guiding centre radius, $R_0$.  {\it Bottom panel:} Perturbation profile,
  $f_{\rm iso}$ (solid curve), and fitting formula given by
  eq.~\ref{eq:fisofit} (dashed curve).  
  \label{figs:contours}}
\end{figure}

A few closed loop orbits are shown in the top panel of
Figure~\ref{figs:contours} along with the isopotential contours of the
perturbed NFW potential.  Also shown are the corresponding isodensity
contours, calculated by solving Poisson's equation.
The major axis of the halo isopotential contours is perpendicular to
the major axis of the closed loop orbits. The bottom panel shows the
eccentricity, $e=1-R_{\rm min}/R_{\rm max}$, of the orbits as well as
that of the isopotential and isodensity contours.  In general, the
isopotential contours are more circular than either the isodensity
contours or the orbits, and all three become more elongated at small
radii, despite the fact that the magnitude of the perturbation
decreases towards the center.

This is shown in the bottom panel of Figure~\ref{figs:contours}, which
shows the perturbation magnitude as function of radius.  The magnitude
of the perturbation $f_{\rm iso}(R)$, increases from zero at the
center to a maximum of $0.375\%$ at $r \simeq 0.25~r_{0.3}$ and is
well fit by a function of the form:

\begin{equation}
f(R) = a~x \exp(-x/b),
\label{eq:fisofit} 
\end{equation}

where $x\equiv R/r_s$ and the best fit parameter values are $a=0.1$ and
$b=0.098$.  We note that the perturbation described by eq.~\ref{eq:fisofit}, or
any similar function which peaks at roughly the same radius and amplitude, gives
similar results as the exact solution for $f_{\rm iso}(R)$.

\section{Simulated Rotation Curves}
\label{sec:simrot}

We now investigate the rotation curves expected when observing a thin
gaseous disk whose kinematics are described by the closed loop orbits
presented in the previous section.  Figure~\ref{figs:rotcur} shows the
orbits for a disk projected at an inclination angle $i=60^\circ$.  The
upper panels of each of the four sets show the projected image of the
disk, color-coded by line-of-sight velocity $V_{\rm los}$. The
isovelocity contours are drawn at $10~\kms$ intervals.  The lower
panels show the rotation curve obtained by sampling the disk with a
thin slit placed along the major axis of the disk, as indicated in the
upper panels by the solid horizontal line.  Also shown are the NFW and
pseudo-isothermal velocity profiles and rotation curve data from one
of the two galaxies in Figure~\ref{figs:rc_allgamma}, scaled to
$(r_{0.3}, V_{0.3})$.

The upper left, upper right, and lower left sets of panels of
Figure~\ref{figs:rotcur} show, for $f=f_{\rm iso}(R)$, the disk when viewed
along various orientations, so that the angle between the projected minor axis
and the minor axis of the orbits is $\phi_a=0^\circ, 15^\circ,$ and
$30^\circ$, respectively.

When the two minor axes coincide, $V_{\rm los}(R) =
V_\phi(R,\phi_0=0^\circ)
\sin(i)$, and the rotation curve matches the $V_{\rm iso}$ curve by
construction (see open circles in the rotation curve panel).  Note that for
this projection, the isovelocity contours of the disk are exactly symmetric,
and the elliptical nature of the orbits is degenerate with the inclination and
circular velocity profile, so that the triaxiality of the halo would be
difficult to detect. Indeed, a slit placed across the minor axis of the
disk---a crude but useful method of assessing triaxiality---would show {\it no
net rotation} for this projection (see diamonds in the rotation curve
panel). This ought to be taken into account when searching for signatures of
triaxiality in $2D$ velocity fields, and it weakens the argument of 
\cite{GENTILE05}, who argue against triaxiality from the lack of strong minor axis
rotation in the HI velocity field of DDO 47. Actually, the isovelocity
contours in DDO 47 (see their Figure 1) show clearly some of the twists
and asymmetries expected for elliptical orbits, as we discuss below.

Under a different projection, $\phi_a=15^\circ$, subtle differences in the
velocity field are noticeable.  Firstly, the isovelocity contours are no
longer symmetric about the $x$ axis.  Instead they exhibit 
``twists'' near the centre of the projected disks, although the velocity field
is symmetric relative to a $180^\circ$ rotation.  As a result, the minor axis
no longer traces an isovelocity contour, and some ``rotation'' is seen along
the minor axis, although confined to the very center.  Note that the shape of
the major axis rotation curve still matches quite well the pseudo-isothermal
profile and the linearly rising rotation curve of UGC~5750.

The lower left panels of Figure~\ref{figs:rotcur} shows the disk from a more
skewed viewing angle, $\phi_a=30^\circ$.  The major axis rotation curve in
this case resembles better the true NFW circular velocity profile and the
rotation curve of UGC~5721. The distortions in the isovelocity contours are
now more pronounced; in particular, the shape of each contour is different in
the upper half from the bottom half of the galaxy but it is still symmetric
relative to a $180^\circ$ rotation. Loosely speaking, one may see the imprint
of the halo triaxiality in a velocity field that differs significantly in
contiguous quadrants, but is anti-symmetric in diagonally opposite ones.

Complex velocity fields requiring more than just circular, coplanar motion in
``tilted rings'' are the norm rather than the exception near the center of
LSBs. This has been clearly demonstrated by the high-resolution combined HI,
H$\alpha$, and CO maps of 5 LSBs presented recently by \cite{SIMON03,
SIMON05}. As these authors conclude, departures from simple circular motion is
clear in the majority of galaxies in their sample. 

Although Simon et al interpret these deviations in term of ``radial
motions'' (i.e., pure expansion or contraction component of the tilted
rings) the interpretation of this complexity in terms, instead, of
elongations in the potential remains viable. In particular, we note
that the isovelocity contours of NGC 2976 (Figure 4a of Simon et al
2003) show clearly some of the twists illustrated in the lower left
panel of Figure~\ref{figs:rotcur}. We also note that their conclusion
about the lack of a cuspy dark matter halo relies heavily on assuming
that the rotational velocity component of their tilted-ring fit traces
the circular velocity of the halo, an assumption that may require
revision if the non-circular motions seen in their data are ascribed
to the effects of triaxiality. Ongoing work is aimed at reaching a
more definitive conclusion on this.

Finally, the lower right panels of Figure~\ref{figs:rotcur} show, for
comparison, the velocity field of a disk in circular motion in an unperturbed
($f=0$) NFW potential, whose rotation curve is in this case indistinguishable
from the long-slit, $f=f_{\rm iso}$, $\phi=30^\circ$ rotation curve. This
illustrates that long-slit rotation curves cannot discriminate between these
two cases; two-dimensional velocity fields are needed.

\section{Discussion}
\label{sec:disc}

Given the variety of rotation curve shapes that arise when considering gaseous
disks in non-axisymmetric halo potentials, the outlook for reconciling dark
matter cusps with LSB rotation curves is promising. In particular, the fact
that departures from circularity are expected to increase towards the center,
and that relatively large effects are possible even in the presence of minor
asphericity appear rather encouraging.  

However, it would be premature to argue that the problem has been
fully solved. Although from the analysis presented here it appears
that, on general grounds, one may expect at least some rotation curves
to be consistent with cored profiles, it is unclear at this point
whether the frequency (and sizes) of apparent ``cores'' is consistent
with this interpretation. This is likely to depend critically on the
triaxiality of the potential in the very inner regions, as well as on
the response of the halo to the assembly of the disk and its
orientation relative to the halo's principal axes. All of these issues
are still quite uncertain.

It is therefore important to build a more compelling case for the triaxial
interpretation of cores in LSB rotation curves, so as to render it
falsifiable. Are the shapes of CDM halos in fact consistent with the perturbed
potentials we have explored?  We are currently investigating this using a sample
of high-resolution galaxy-sized halos.  Are there any corroborating traits that
may be used to confirm or exclude the hypothesis that halos surrounding LSBs are
indeed triaxial?  Baryonic matter in the form of a bar could have a similar
effect on gas motions as a triaxial dark matter halo \citep{RHEE04}.  

How can one best verify the triaxial halo interpretation in
two-dimensional velocity maps?  Identifying a clean and unambiguous
indication of triaxiality, such as the unusual isovelocity ``twists''
and symmetry patterns shown in Figure~\ref{figs:rotcur}, will be as
important as the success of aspherical halos in reproducing the rich
variety of shapes of LSB rotation curves.  Only if this is
accomplished shall we be able to conclude that LSB rotation curves do
not preclude the presence of dark matter density cusps, thereby
freeing the CDM paradigm of one vexing challenge on small scales.

\section*{Acknowledgements}
This work has been supported by various grants to JFN from NSERC, CFI, and by
fellowships from the Alexander von Humboldt Foundation.  We thank the members of
our long-term N-body collaboration, Carlos Frenk, Simon White, Adrian Jenkins,
Volker Springel, and Chris Power for valuable discussions during the course of
this project. Joachim Stadel, Tom Quinn, and James Wadsley are also thanked for
giving us access to their excellent hydrodynamic code GASOLINE, which we used in
the early stages of this project. We also thank Josh Simon and Alberto Bolatto
for useful discussions.  We thank Linda Sparke for valuable insights that
greatly improved this work.  We thank the anonymous referee for many useful
comments and suggestions which significantly improved this manuscript.

\bibliography{stan}

\onecolumn
\begin{figure*} 
\plotfour{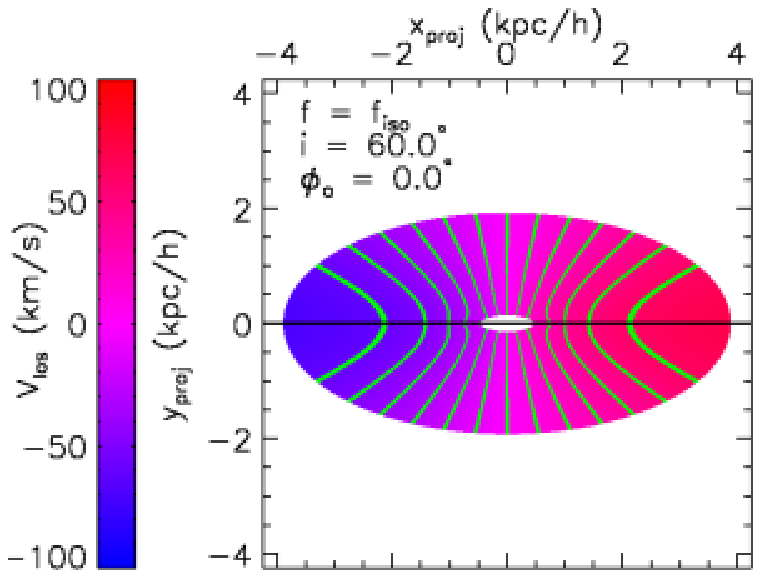}{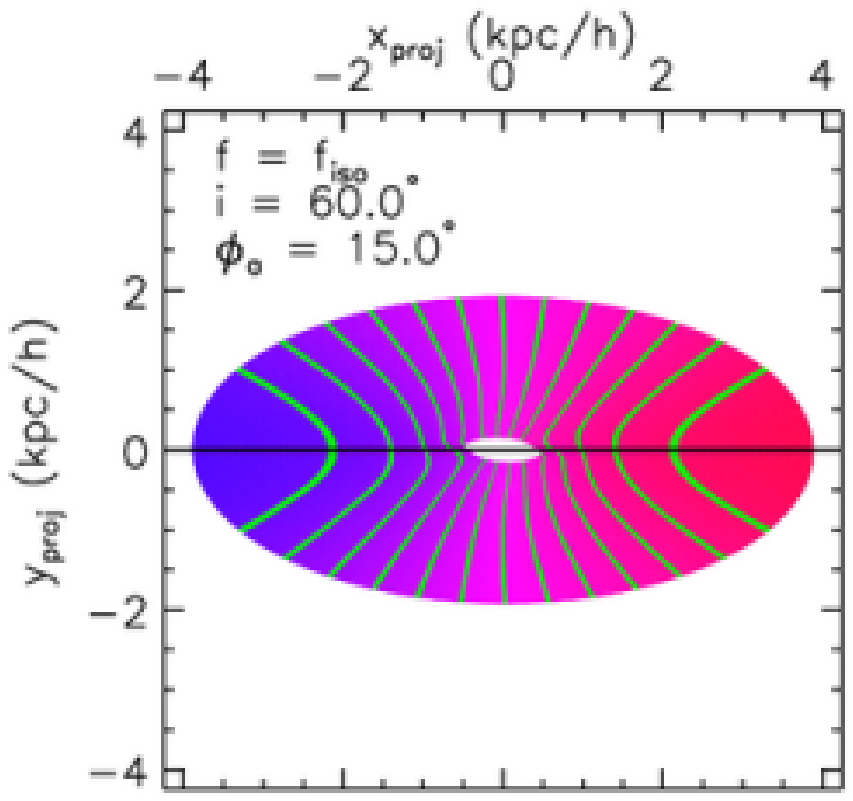}{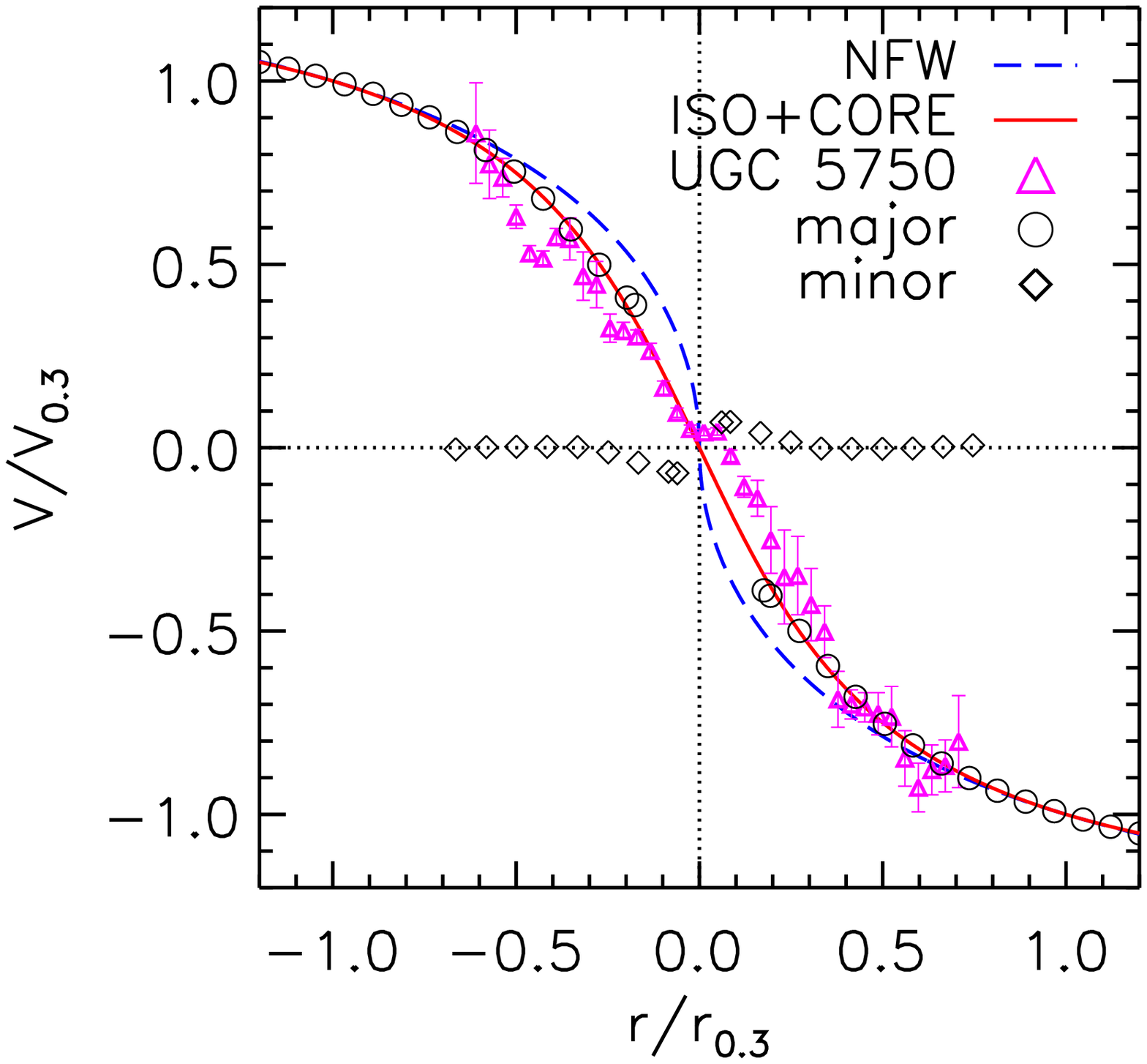}{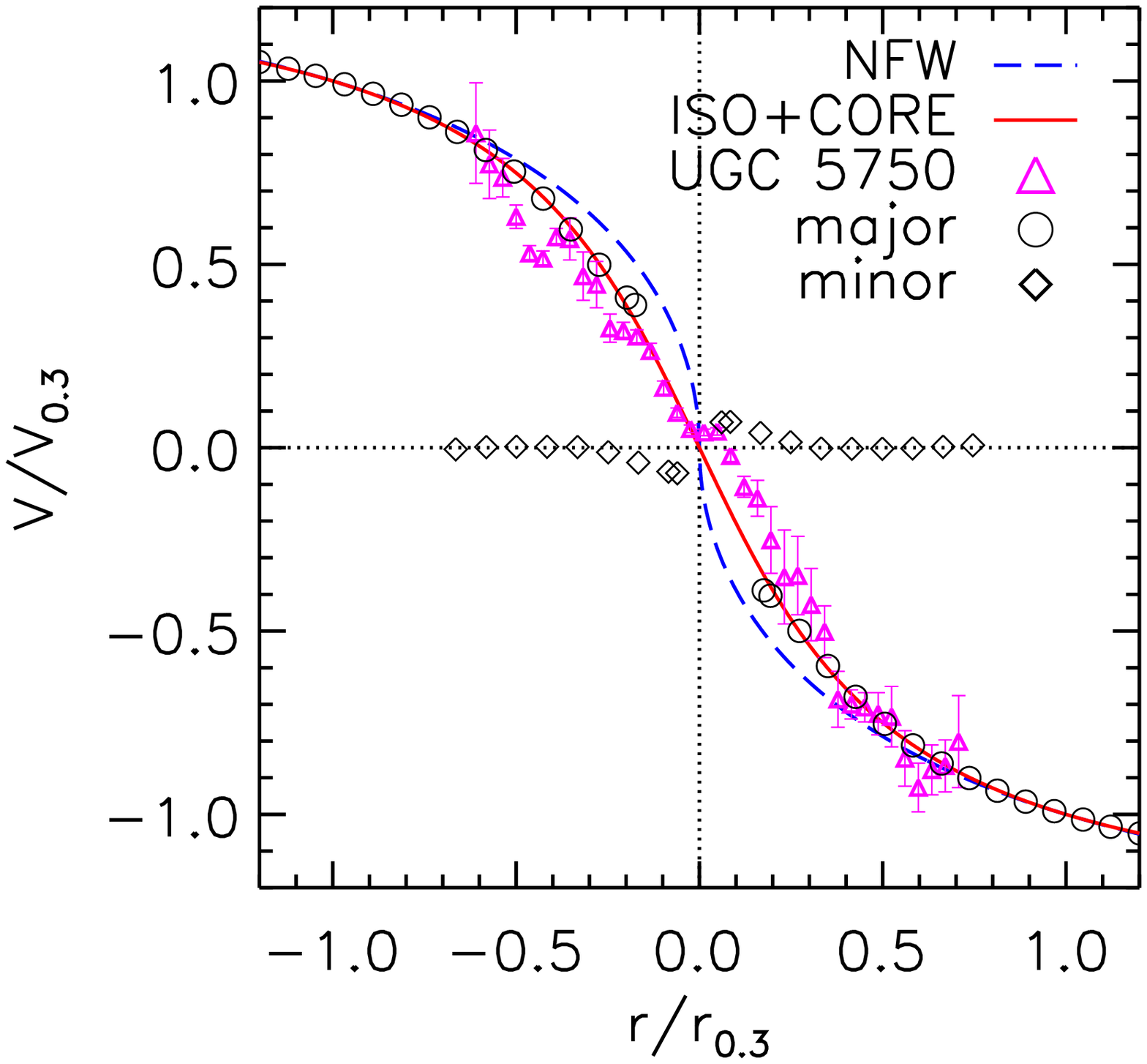}
\end{figure*}

\begin{figure} 
\plotfour{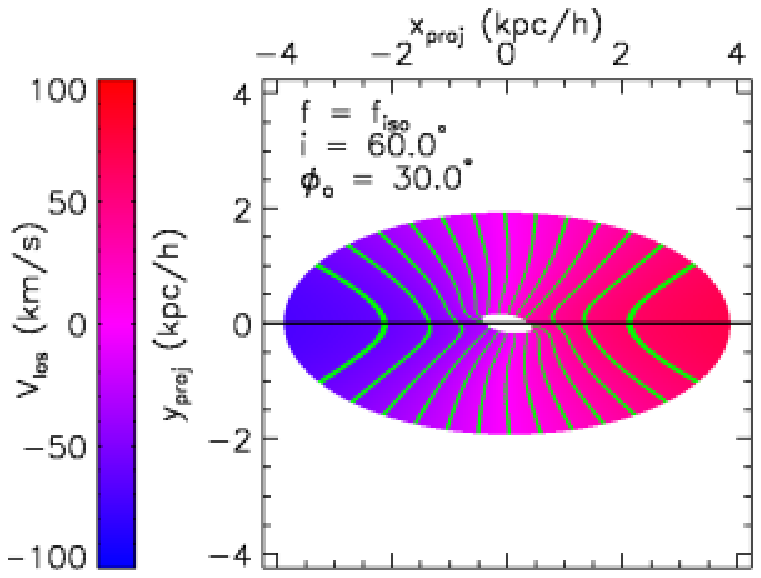}{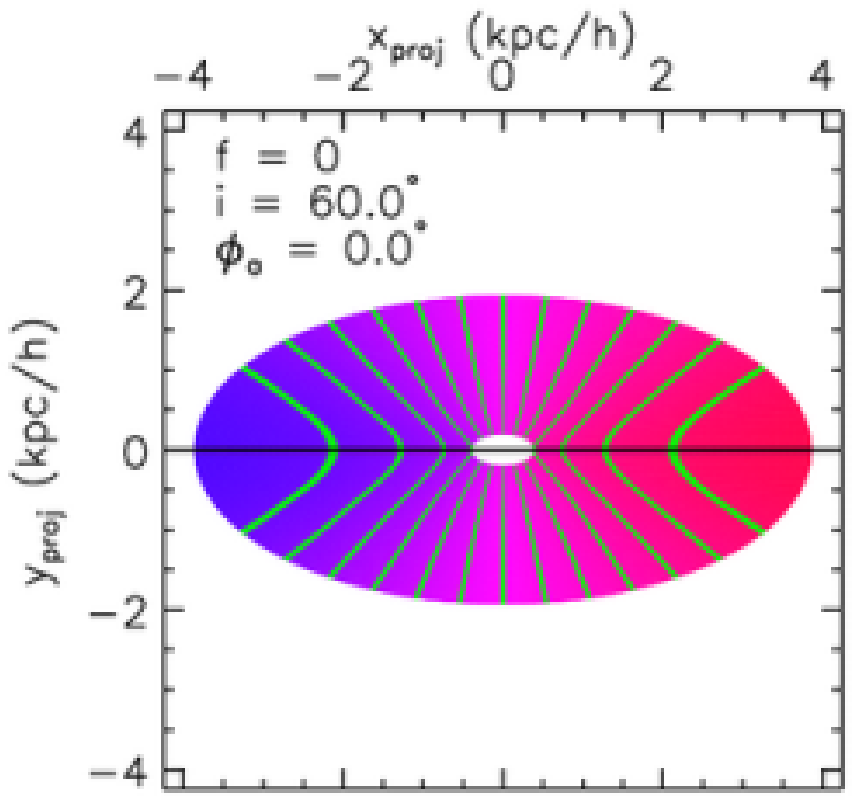}{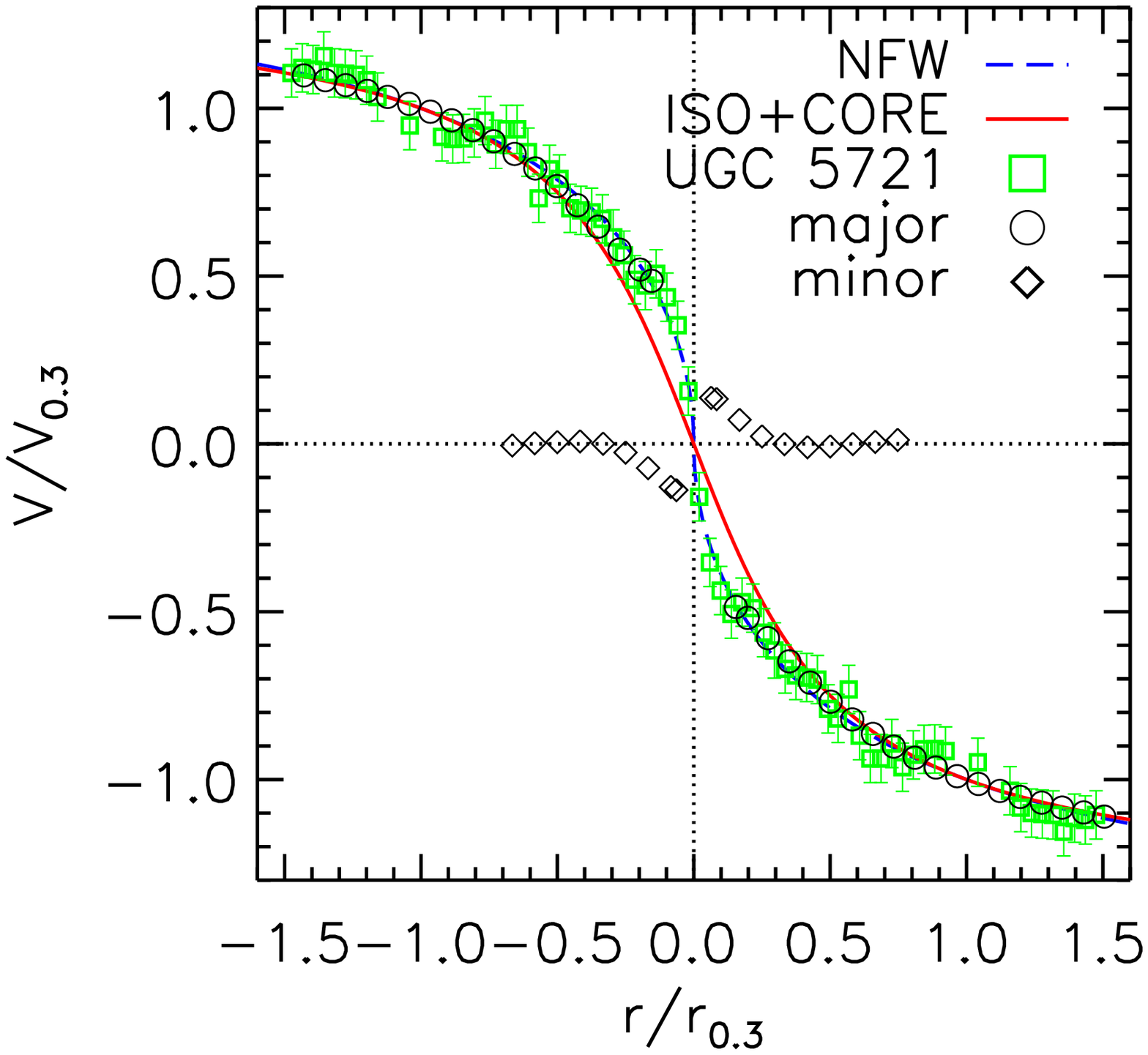}{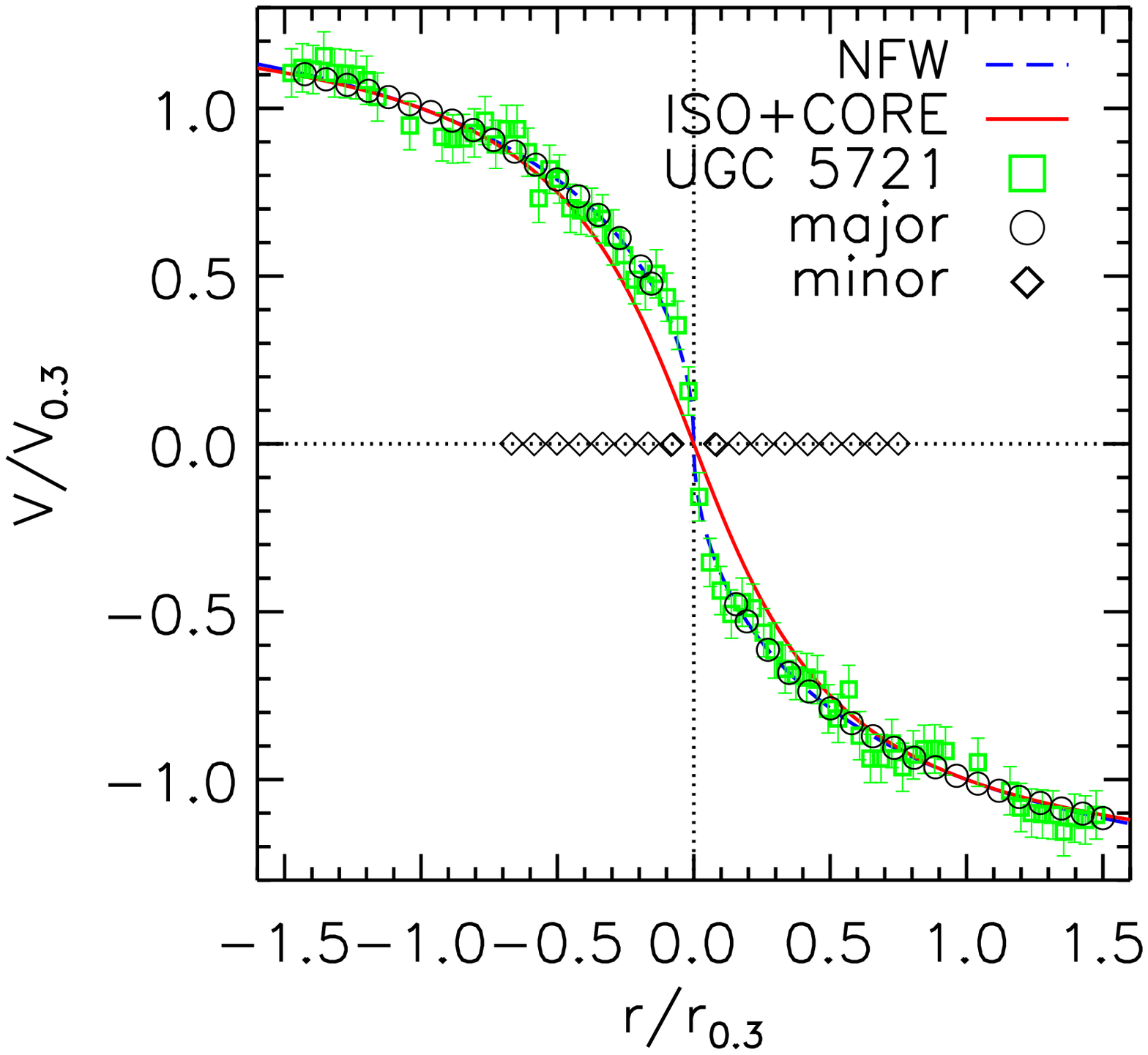}
\caption[Rotation curve of disk as inferred from simulated long-slit radial
  velocity data] {{\it Upper panels of each set:} Projected velocity fields of
  disks in perturbed $(f=f_{\rm iso})$ and unperturbed $(f=0)$ NFW potentials,
  inclined by $i = 60^\circ$.  The disks are color-coded by line-of-sight
  velocity as indicated by the color bar.  The green curves show isovelocity
  contours in $10~{\rm km/s}$ intervals.  The solid black line indicates the
  position of a long slit placed along the major axis of the projected disk.
  The $(f=f_{\rm iso})$ disk is shown for three different values of $\phi_a$, the angle between the slit and
  the long axis of the closed loop orbits.  {\it Lower panels of each set:} Rotation curves corresponding
  to the disks in the upper panels, scaled to $(r_{0.3}$, $V_{0.3})$. Open
  circles (open diamonds) show the velocity profile along the slit placed
  across the major (minor) photometric axis of the projected disk.  Open
  triangles (squares) show the rotation curve of UGC~5750 (UGC~5721) from
  Figure~\ref{figs:rc_allgamma}.  The shape of the rotation curve of the
  $f=f_{\rm iso}$ disk agrees with that of UGC~5750, which rises linearly with
  radius for $\phi_a=0^\circ$ and $15^\circ$, and is similar to the
  pseudo-isothermal velocity curve shown by the solid curve.  The minor axis
  velocities are zero when the disk is viewed at $\phi_a=0$, but
  nonzero minor axis velocities and twisted isovelocity contours result when a
  different line-of-sight, $\phi_a=15^\circ$ or $30^\circ$ is chosen.  The
  $f=f_{\rm iso}$, $\phi=30^\circ$ and $f=0$ disks have major axis rotation
  curve shapes similar to or identical to the NFW velocity profile, and match
  closely the UGC~5721 rotation curve.  The velocity field of the disk in the
  perturbed potential can be distinguished from the unperturbed NFW by the
  nonzero minor axis velocities and the twisted isovelocity contours.
\label{figs:rotcur} }
\end{figure}

\end{document}